\documentclass[12pt,preprint]{aastex}

\shorttitle{Double nucleus in M83} \shortauthors{Mast et al.}

\begin{document}
\title{Double nucleus in M83} \author{Dami\'an Mast\altaffilmark{1}, Rub\'en J.
D\'{\i}az\altaffilmark{1,2}, M.Paz Ag\"uero \altaffilmark{1}}

\altaffiltext{1}{Observatorio Astron\'omico de C\'ordoba y
CONICET, Universidad Nacional de C\'ordoba, Laprida 854, 5000
C\'ordoba, Argentina.} \altaffiltext{2}{Gemini Observatory,
Southern Operations Center, c/o AURA, La Serena, Chile.}

\email{damian@mail.oac.uncor.edu}

\begin{abstract}

M 83 is one of the nearest galaxies with an enhanced nuclear star
formation and it presents one of the best opportunities to study
the kinematics and physical properties of a circumnuclear
starburst. Our three-dimensional spectroscopy data in $R$ band
confirm the presence of a secondary nucleus or mass concentration
(previously suggested by Thatte and coworkers). We determine the
position of this hidden nucleus, which would be more massive than
the visible one, and was not detected in the optical HST images
due, probably, to the strong dust extinction. Using a Keplerian
approximation, we estimated  for the optical nucleus a mass of
$(5.0\pm0.8)\times10^6$M$_{\odot}/ \sin i\,\, (r < 1\farcs5)$, and
for the hidden nucleus, located $4\arcsec\pm1\arcsec$ at the NW
(PA $271\degr\pm15\degr$) of the optical nucleus, a mass of
$(1.00\pm0.08)\times10^7$M$_{\odot}/ \sin i\,\, (r < 1\farcs5)$.
The emission line ratio map also unveils the presence of a second
circumnuclear ring structure, previously discovered by IR imaging
(Elmegreen and coworkers). The data allow us to resolve the
behavior of the interstellar medium inside the circumnuclear ring
and around the binary mass concentration.

\end{abstract}
\keywords{galaxies: individual (M83) —-- galaxies: kinematics and
dynamics —-- galaxies: nuclei —-- galaxies: starburst —--
galaxies: structure —-- techniques: spectroscopic}

\section{INTRODUCTION}

M 83 is a very special case of nearby grand design barred spiral
galaxy (Table 1), it harbors a nuclear X-ray source \citep
{soria02}, coexisting with a massive starburst in a double
circumnuclear ring \citep {elmegreen98} and a secondary mass
concentration recently suggested by Thatte et al. (2000) and
confirmed by 3D spectroscopic data \cite {mast02}. Gallais et al.
(1991) obtained infrared (IR) images of the central 20\arcsec\ in
the  $J$, $ H $ and $K$ bands and analyzed the color-color
diagrams which indicate that the nucleus and several regions in an
arc at 7\arcsec\ from the nucleus are forming stars at a high
rate. HST observations (optical and $UV$) and IUE spectra reveal
that the arc contains more than 20 massive young clusters similar
to 30 Dor in the Large Magellanic Cloud (LMC), with ages between 2
and 8 million years and masses in the range
$1\times10^4-1\times10^6$M$_{\odot}$ \citep {heap93, heap94}.

Thatte et al. (2000) made $IR$ spectroscopic observations with two
long slit positions of the nuclear region using the ISAAC
spectrometer at the VLT. They detected two peaks in the radial
velocity dispersion profile, one on the $K$-band luminosity peak,
and the other one displaced about 3\arcsec\ to the SW, suggesting
the presence of a secondary nucleus. The lack of two spatial
dimensions in the spectroscopic information did not allow them to
fix the precise position of the second mass concentration.
Although the morphology of the star formation activity has been
studied in detail \citep {gallais91, harris01}, there are very few
kinematic or dynamical studies based in good spatial samplings. In
this work we present a study of the 2D kinematics of the central
region of this galaxy, which resolves the behavior of the
interstellar medium around the binary mass concentration.

\section{OBSERVATIONS}

A summary of the observations is presented in Table 2. We made
three-dimensional (3D) spectroscopic observations of the central
region of M 83 using the Multifunctional Integral Field
Spectrograph mounted on the 1.54 m telescope at Bosque Alegre
Astrophysical Station \cite [2003]{diaz99}. The data were gathered
between March 2001 and May 2002. Several circumnuclear fields were
observed, we present here the analysis of the selected field that
includes both mass concentrations.

The analyzed integral field was obtained with a total exposure
time of 2 hours. The spectra, with a resolution of 1.5\,\AA\ in
the $R$ band, were obtained with an array of $8 \times 14$ lenses
(1\farcs5 field each one) sampling a field of $12\arcsec\times
21\arcsec$. The seeing was about $2\arcsec$ so the field is
slightly subsampled.  We complemented these data
with archival HST images in the filters F187N, F190N (Pa$\alpha$
and Pa$\alpha$ continuum), F814W (7940\,\AA\,) and F656N
(6563\,\AA\,).

\section{DATA REDUCTION AND ANALYSIS}

The reduction of the spectra was performed with the software SAO
(developed by the Special Astrophysical Observatory, Russia), the
software ADHOC \citep{boulesteix93} and PC standard worksheets.
The procedure followed is described in detail in the works of
D\'{\i}az et al. (1997), Vega (2000) and Mast (2000). The spectra
include the nebular emission lines H$\alpha$, [NII] 6548\,\AA\,,
6583\,\AA\,, [SII] 6716\,\AA\ and 6731\,\AA\, which were fitted
using gaussian curves to determine the parameters that describe
them. In most of the field, the $S/N$ allowed the detection and
fitting of the mentioned emission lines, but in some locations (in
the field SW border) the $S/N$ ratio of the spectrum was low and
only H$\alpha$ and [NII] 6583\,\AA\, emission were detected. We
made gaussian fitting up to a level of $20\%$ of the peak value.
Therein the profile wing asymmetries that in some locations could
appear do not affect strongly the manual interactive fitting
performed to the lines. Once the emission line gaussian fitting
was completed, we obtained the following data for each lens:
H$\alpha$, [NII] 6583\,\AA\,, [SII] 6717\,\AA\, and [SII]
6731\,\AA\, fluxes, wavelength positions and FWHM of each line.
With this information the following maps were constructed: Red
continuum intensity, H$\alpha$ intensity, [NII] 6583/H$\alpha$,
([SII] 6717+[SII] 6731)/H$\alpha$, [SII] 6717/[SII] 6731, and FWHM
maps. Because the H$\alpha$ line is the most conspicuous emission
of the obtained spectra and the one showing better $S/N$ ratio, we
only analyzed the radial velocities corresponding to this line. A
map of the radial velocity field was also constructed.

As mentioned before, the field was slightly subsampled. In order
to improve the data processing and visualization, all the field
values were interpolated, yielding  the same field sampled by $24
\times42$ elements (0.5\arcsec/pixel). The internal error due to
the velocity determination procedure has a relative value lower
than 5 \,km\,sec$^{-1}$ (r.m.s.). That is the precision limit in
the velocity field determination, in case we have data with good
$S/N$ ratio ($>10$). In our case the best $S/N$ ratio in the
H$\alpha$ emission line is $S_{Total}/N = 30$, $S_{Peak}/N = 15$;
the worst $S/N$, in one of the outermost lenses, is $S_{Peak}/N =
2$.

\section{RESULTS}

\subsection{Morphological scenario}

Figure 1 shows the central $40\arcsec\times40\arcsec$ around the
optical nucleus from which a bright small red arc emerges to the
SW, but the most conspicuous feature is a giant star forming arc
\citep {gallais91, elmegreen98}. In this one, to the NE of the
nucleus several star clusters are present and have been studied by
Harris et al. (2001). Going counterclockwise through the arc, a
highly obscured region is located with a large number of dust
patches and a star forming region emerging from the dust, more
than 5\arcsec\,westward from the nucleus. The $R$ band continuum
map (Figure 2) shows an ellipsoidal light distribution with the
external isophotes center, determined from the red continuum HST
image (F814W filter), located 2\farcs5\,to the W-SW of the
 continuum emission peak (PA $230\degr$), which we assumed, corresponds to the optical
nucleus position. This result is consistent with the observations
of Thatte et al. (2000), who found that the nucleus is not located
at the symmetry center (of the circumnuclear light distribution)
in their $K$ band images. The symmetry center of the outer
isophotes in $R$ and $K$ bands, which could be considered the
bulge center, is almost the same, implying that the optical
nucleus would be not located at the stellar bulge center (see the
sketch in Figure 3).

\subsection{Radial Velocity Field of the Ionized Gas}

Figure 4 shows the obtained velocity field, which covers
$12\arcsec\times21\arcsec$ of the central region of M\, 83.  A
square marks the position of the optical nucleus, which
corresponds to the continuum image peak (at $R$ band, $N_{R}$ in
Figure 3 and 4).  The isovelocity lines were traced each
10\,km\,sec$^{-1}$ (2 $\sigma$ uncertainty). The isovelocity lines
are strongly distorted and could indicate a departure from the
axisymmetric potential, with a main line crowding at the position
indicated by a circle, where the hidden nucleus should be located.
The radial velocity field has a distortion that indicates the
presence of two mass concentrations, neither of them located
exactly on the global velocity field minor axis (dotted line in
Figure 4 and 7).

The radial velocity field is similar to that of the central region
of NGC\,3227 (Arribas \& Mediavilla 1994) which shows an AGN out
of the kinematic center of the galaxy. A previous clear case of a
kinematic center not located at the bar+bulge center of symmetry
was found by D\'{\i}az et al. (1999) in the 2D spectroscopic maps
and images of NGC 1672, a southern barred galaxy with a strong
circumnuclear starburst.

The extreme velocity values of the observed field are $V_{max}$=
634\,km\,sec$^{-1}$ and $V_{min}$= 542\,km\,sec$^{-1}$($\Delta$V=
92 \,km\,sec$^{-1}$).

There is an isovelocity line stretching (Figure 5) which is
coincident, within the uncertainties, with the continuum peak
previously identified with the optical nucleus, and therefore is
about 2\farcs5\, to the N-NE of the red continuum symmetry center
($C_{R}$ in Figure 3 and 4).

The radial velocity gradient determined by the line stretching
indicates that the mass concentration of the hidden nucleus would
be larger than that of the visible one.

Areas with predominant circular motion can be drawn around each
mass concentration (see Figure 6). Each area involves 6 lenses
(optical nucleus) and 11 lenses (hidden nucleus).

The mass concentration corresponding to the optical nucleus and,
within the uncertainties, coincident with the continuum peak
position, is at $2\farcs3\pm0\farcs5$ to the W-SW from the red
continuum symmetry center, while the hidden nucleus, kinematically
determined, would be at $2\farcs4\pm0\farcs5$ NW from the red
continuum symmetry center. Therefore, the hidden mass
concentration is $3\farcs9\pm0\farcs5$ at PA $271\degr\pm15\degr$,
from the optical nucleus (continuum emission peak). Using the
astrometric parameters of the HST WFPC observations, we derive the
position of the main nucleus: (J2000) $\alpha$ = 13h 37m 00.919s,
$\delta$ = $-29\degr 51\arcmin 55\farcs66$ with a $0\farcs1$
uncertainty. Assuming that this is the position of the peak
continuum emission in our maps, the coordinates of the hidden
nucleus would be $\alpha$ = 13h 37m 00.540s $\pm\,0.04$s, $\delta$
= $-29\degr 51\arcmin 53\farcs62 \pm 0\farcs5$, considering an
uncertainty in the position of $1/3$ of sampling element. The
position was determined from cosine fitting to the radial
velocities azimuthal profiles. The uncertainty is determined by
the amount of rotation center displacement that makes the fit
clearly wrong. This displacement is about $d/\sqrt{n}$  where $d$
is the sampling element size in arcseconds and $n$ is the number
of sampling elements that show a rotation pattern. The hidden
nucleus is at the NW extreme of the star formation arc. Its
position can be projected on the long slit direction of Thatte et
al. (2000), and this projection is at the same position were the
authors note a local maximum in the stars velocity dispersion. The
main difference is $2\farcs5$ along a direction perpendicular to
the slit of Thatte el al (2000).

The observed region is too small on galactic scale, also perturbed
and, in order to derive the kinematic parameters, we smoothed the
observed radial velocity field with a median filter of size
$6\arcsec \times 6\arcsec$. The result is shown in Figure 7. The
PA of the minor axis is 138\degr, very near to the PA of the low
resolution CO velocity field determined by Lundgren et al. (2004)
of $136\degr$. The masses can be estimated from the Keplerian
approximation
\begin{equation}\label{1}
    \textrm{M}   \sim  233 \,\, \textrm{V}^{2} \, \textrm{R} \, /\sin
    i,
\end{equation}

where $i$ is the inclination, V the velocity in km sec$^{-1}$), R
de radius in pc and M the mass in M$_{\odot}$. Therefore we
assumed for the masses estimations, an inclination equal to the
global inclination value of 24\degr\,\citep{comte81}. The
estimated masses of the nuclei within a 1\farcs5 radius would be
around $(1.2\pm0.2)\times10^{7}\, $M$_{\odot}$ (optical nucleus)
and $(2.4\pm0.2)\times10^{7}\,$M$_{\odot}$ (hidden nucleus). There
is no intention to report strict Keplerian rotation around the
mass concentrations, we just note a predominance of circular
motions involving several detection elements (lenses) and we
derive upper limits for the masses in a Keplerian approximation.

In order to see if the gas is in ordered rotation around the mass
concentrations we studied the line profiles: a strongly asymmetric
line profile, with pronounced wings on one side is more likely due
to non gravitational motions. As can be seen in Figure 8, this is
not the case for the spectra in the hidden nucleus position.

\subsection{FWHM map}

Figure 6 shows a FWHM map with the radial velocity field
superposed. Two local peaks coincident, within the uncertainties,
with the position of the two mass concentrations indicated in
Figure 4, can be seen.  We note here that the lack of exact
coincidence between kinematic centers and FWHM peaks, could be
real. This is the case of the eccentric disk of M 31 (See Figure 7
of Bacon et al. 2001) where the lack of symmetry in the
gravitational potential generates a difference in the locations of
the rotation center and the observed velocity dispersion peaks at
a given resolution. It is expected that stellar rotation (not gas)
around the mass center is not necessary circular because it is
subject to perturbations (like the presence of a nearby second
nucleus). In this case, the radial velocity dispersion peak is not
coincident with the rotation center, due to the ellipticity in the
orbits \citep{bacon01, tremaine95}. Therefore, the motion is
circular only in an approximate approach which allows the mass
estimations.

In the SW part of our FWHM map, profile widths exceed 200 \,km\,sec$^{-1}$.
This could be due to shocks in the ionized
gas considering that this region coincides with the star forming
arc.

\subsection{[NII] 6583 / H$\alpha$ map}

Figure 9 shows the [NII] 6583 / H$\alpha$ ratio. The position of
the optical nucleus is marked with a star. A ring of mean value
0.62 is appreciable around the optical nucleus, coincident with
the inner ring reported by Elmegreen et al. (1998) (Figure 10).
The ratio is comparatively low in the inner ring region (nuclear
region) and in the star forming arc, ranging from values between
0.40 in the star forming arc to 0.46 in the inner ring region.
Those values are the expected for normal HII regions (lower than
0.5, Osterbrock 1987). The peak value in the ring is $\sim 0.73
\pm 0.02$. Assuming that those ratios depend mainly on the
abundance effects, their distribution would indicate that the
relative abundance of nitrogen is higher in the ring than in the
nuclear region and the star forming arc. The low I($\lambda$6584)
in relation to H$\alpha$ would suggests that the starburst in the
nuclear region and in the star forming arc is younger than in the
ring. The values at the region of 3\arcsec\, around the hidden
nucleus are $\sim 0.66 \pm 0.02$.

\section{DISCUSSION}

It is important to note that the circular motion around both
regions has been measured within several detection elements.
Furthermore the local kinematics suggests that the secondary
nucleus is more massive, but another possibility must be
mentioned: a warp in the nuclear region could eventually mimic
strong distortions of the kinematic minor axis, but the shape of
the observed velocity field with two isoline stretchings would be
very difficult to reproduce with a mild local change of
geometrical parameters and a single warp model. Other
possibilities were considered, like the kinematic distortions
arisen in bars and spiral arms, but: i) the shape of the radial
velocity field cannot be successfully compared with those
resulting from numerical modelling like those by Piner et al.
(1995) for a barred potential; ii) we do not detect at $K$ band,
nuclear bars or spiral arms in the region of the velocity field
distortion.
 In order to yield much greater insight into the
physics of the nuclear region, we use the H$\alpha$ and Pa$\alpha$
HST images, together with case B recombination theory, to
construct an extinction map of the line emission from the
H$\alpha$/Pa$\alpha$ ratio (Figure 11). This was made in the same
way that Harris et al (2001) used the  H$\alpha$/H$\beta$ ratio.
Using a two-component model for the dust extinction and following
Calzetti (1997), the standard definition of the color excess is
\begin{equation}\label{2}
     \textrm{E}(B-V)_{H_{a}/H_{b}}=\frac{\textrm{log}(R_{obs}/R_{int})}{0.4[\kappa(\lambda_{a})-\kappa(\lambda_{b})]},
\end{equation}

where $R_{int}$ and $R_{obs}$ are the intrinsic and observed
hydrogen line ratios, respectively, and $\kappa(\lambda)$ is the
extinction curve, measured at the wavelength of the emission line.
Using the average LMC curve and a Seaton Galactic curve \citep
{fitzpatrick86}, and adopting  R$_{int} = 0.1$ \citep {lilly87,
osterbrock87}, we can derive the attenuation in magnitudes from
\begin{equation}\label{3}
    A_{547} = 1.79[\textrm{E}(B-V)- 0.06] + 3.10 \times 0.06 \approx
    A_{V},
\end{equation}

(from eq.2 of Harris et al. (2001)) where it was assumed a
foreground Milky Way component E$(B-V)_{MW}= 0.06$\,\,\
\citep{schlegel98}. According to our extinction map, the values of
A$_{V}$ range from 0 to nearly 7 magnitudes at the region marked
in Figure 11.

It can be seen that the strong extinction is very localized and
would not play an important role in the kinematics. The position
of the hidden nucleus is, within the uncertainties, coincident
with a knot appearing at $K$ band (hot spot labelled ``region 8"
in Figure 8 of Elmegreen et al. (1998)). This could be the IR
source corresponding to the hidden mass concentration.

The H$\alpha$ map, not shown here, does not clearly reveal the
existence of the innermost circumnuclear ring reported by
Elmegreen et al. (1998). Frequently such rings are constituted by
a discrete series of emission regions. In this case, the ring is
conspicuous in the near IR color maps presented by Elmegreen et
al. (1998) and in our NII/H$\alpha$ map, probably suggesting
different properties (mainly age) in the HII regions, instead of
an enhanced star formation outside this innermost ring. The most
evident H$\alpha$ structure is a star forming arc that extends
from the SE to the NW, which would be part of an outermost
circumnuclear ring \cite {elmegreen98}. The study of abundances in
HII regions in the disks of spiral galaxies has shown the
existence of negative gradients with a higher abundance, with the
consequence of a higher NII/H$\alpha$ ratio (e.g. Pagel \& Edmunds
1981, Evans 1986, Shield 1990). The observed ratios distribution
in M 83  could be another example of this phenomenon. Observations
of the global trend and the differences between the arms, inter
arms and rings regions of the barred galaxy NGC 1566 (Ag\"{u}ero
et al. 2004, Roy \& Walsh 1986), were interpreted as possibly
arisen in different ages of the star forming regions.

One question arises here: if the velocity field distribution
corresponds to a binary mass concentration, can the star formation
be triggered by the passage of the hidden nucleus trough the high
gas density medium? It has been claimed that if a merging
satellite galaxy has no nucleus (e.g., Magellanic Clouds), the gas
in the satellite will interact with the gas in the host disk and
then be settled in the disk before reaching the nuclear region. On
the other hand, if it has a nucleus (e.g., M32), the satellite
nucleus will sink toward the nuclear region because of the
dynamical friction (Taniguchi \& Wada 1996). It is assumed that a
nucleus is either an SMBH or a significant concentration of a
nuclear star cluster. In this respect, satellite galaxies in Mihos
\& Hernquist (1994) and Hernquist \& Mihos (1995) are also
nucleated ones. The formation mechanism of the corresponding
nuclear starburst and hot-spot nucleus would involve a
supermassive binary of compact objects formed by a merger with a
nucleated satellite galaxy triggering intense star formation in
the central regions of the main spiral galaxy, in which the
nuclear gas disk has been formed already by the dynamical effect
of the merger itself. As the secondary compact object approaches
the nuclear gas disk, the gas disk response to the gravitational
perturbation caused by the nonaxisymmetric potential of the binary
mass, forming asymmetrical spiral patterns. When the mass of the
intruding object is only one-tenth of the primary's, the gas
response is so mild that only pseudo-ring features or tightly
wound spiral arms are formed in the circumnuclear region. These
features are expected to evolve into several H II region clumps,
leading to the formation of not very bright hot-spot nuclei. On
the other hand, when the mass of the intruding nucleus is
comparable to half the primary one, a very strong one-arm spiral
shock appears after the close passage of the secondary in each
orbital period. The gas clouds are forced to move rapidly into the
central region owing to frequent collisions of gas clouds and the
starburst is triggered near the nucleus. If this would be the
case, then the age gradient in the star-forming arc (see \S 1)
that ends in the hidden nucleus would provide the clue to the
dynamical evolution of this  complex system.

\section{CONCLUSIONS}

In this work we studied the morphology, the ionized gas emission
and the kinematics in the innermost regions of M 83. A hidden
nucleus is not evident in the HST images, although kinematically
the radial velocity field points out the presence of a strong mass
concentration that could be the mentioned nucleus. Its condition
of non detectable source could be due to its location, considering
that the kinematically derived position lies close to a strong
absorption region ($A_V \sim $7 mag, see Figure 11).

[NII] emission is relatively lower compared with H$\alpha$
emission of the hidden nucleus region, at what could be the
younger end of the starforming arc, according to previous works.
This would be consistent with lower metalicity values at this
region, but the lack of more emission line ratios precludes a
direct confirmation with the data presented here. Notwithstanding,
two independent works confirmed through photometry
\citep{harris01} the age gradient of
 this star forming arc. The emission line ratio map also unveils
the presence of a second circumnuclear ring structure, previously
known by IR imaging \citep {elmegreen98}.

Using the Keplerian approximation, the optical nucleus would have
a mass of $(5.0\pm0.8) \times10^6$M$_{\odot}/\sin i\,\, (r <
1\farcs5)$, and the secondary nucleus, located
($4\arcsec\pm1\arcsec$) at the NW (PA $271\degr\pm15\degr$) of the
optical nucleus, would have mass of
$(1.00\pm0.08)\times10^7$M$_{\odot}/\sin i\,\, (r < 1\farcs5)$.
The FWHM map presents local maxima in the nuclei positions,
supporting the conclusion about the presence of two central mass
concentrations derived from the radial velocity map.

The behavior of the interstellar medium inside the circumnuclear
ring and around the binary mass concentration has been resolved.
The hidden nucleus is located on the younger end of the giant arc
of star formation; suggesting that the local departure of an
axisymmetric gravitational potential would be the trigger of the
nuclear starburst in M 83.

\section{Acknowledgements}

We thank the generous support of Germ\'an Gimeno, Walter Weidmann
and Gustavo Carranza during the observations. We also thank the
anonymous referee, whose comments allowed us to improve the paper
presentation. This research was partially supported by the CONICET
(grant PIP 5697), by the Agencia C\'ordoba Ciencia, and by the
Gemini Observatory, which is operated by the Association of
Universities for Research in Astronomy, Inc., on behalf of the
international Gemini partnership of Argentina, Australia, Brazil,
Canada, Chile, the United Kingdom, and the United States of
America. Some of the data presented in this paper were obtained
from the Multimission Archive at the Space Telescope Science
Institute (MAST). STScI is operated by the Association of
Universities for Research in Astronomy, Inc., under NASA contract
NAS5-26555. Support for MAST for non- HST data is provided by the
NASA Office of Space Science via grant NAG5-7584 and by other
grants and contracts.

\clearpage

% FIGURES

\begin{figure}
\figurenum{1} \epsscale{1}
 \plotone{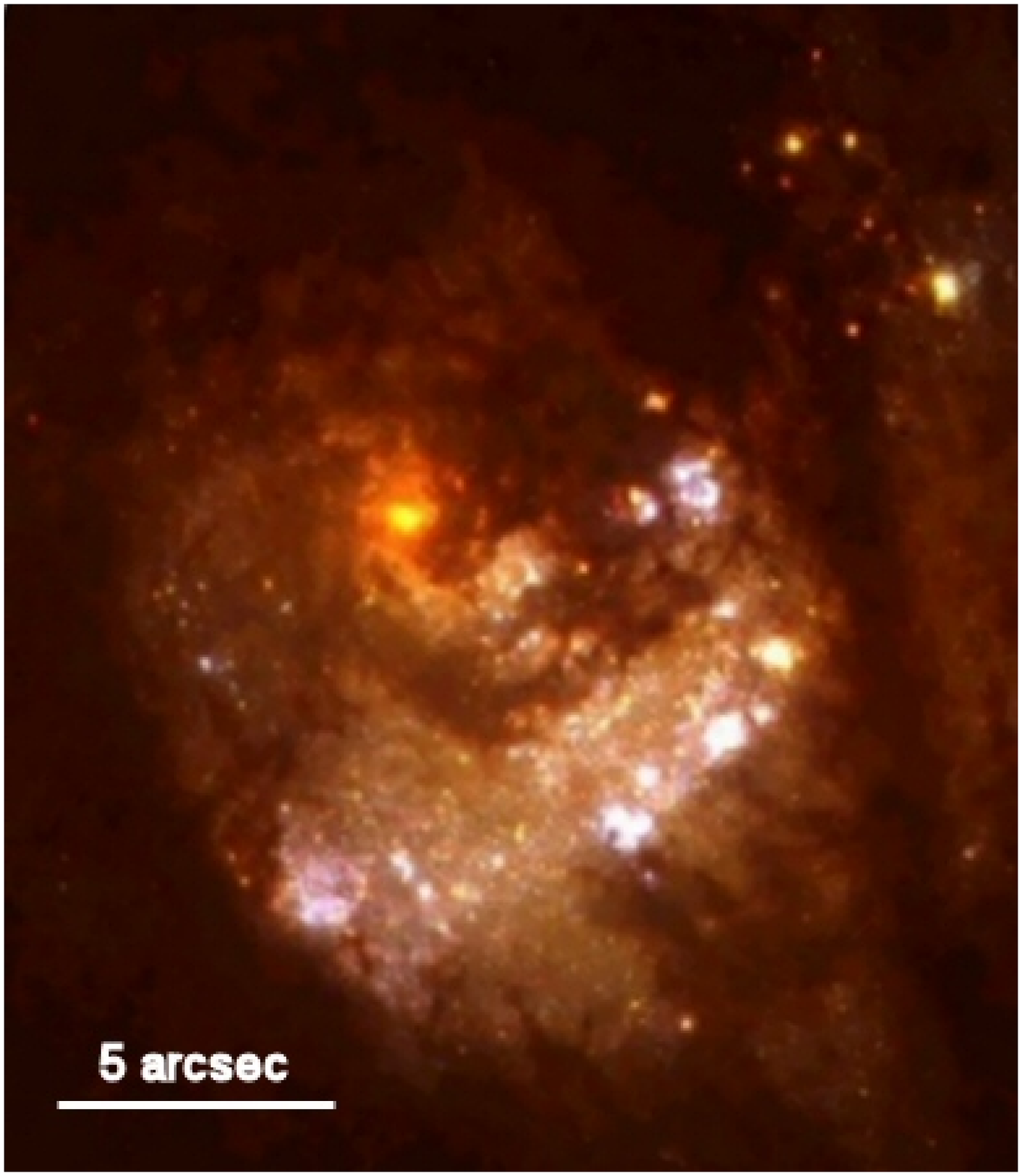}
 \caption{HST false color image combined from F439W (contributing to the white-bluest features in the representation),
 F555W (yellow) and F702W (contributing to the reddest features of the depiction). Point spread functions were matched to a common resolution of
 $0\farcs09$. North is at the top and east is at the
left.}
\end{figure}

\begin{figure}
\figurenum{2} \epsscale{1}
 \plotone{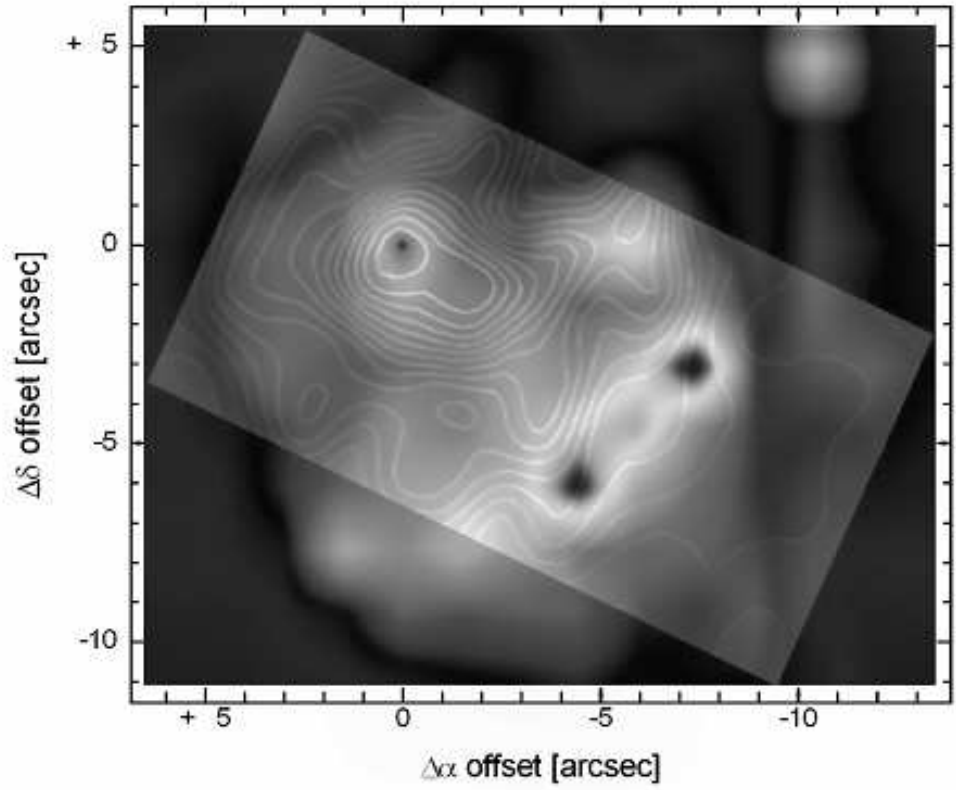}
 \caption{Map of red continuum relative intensity (isophotes)
 computed in the range $6454-6524$ \AA,
superposed to F814W HST image smoothed to 1\arcsec \,seeing.}
\end{figure}

\begin{figure}
\figurenum{3} \epsscale{1}
 \plotone{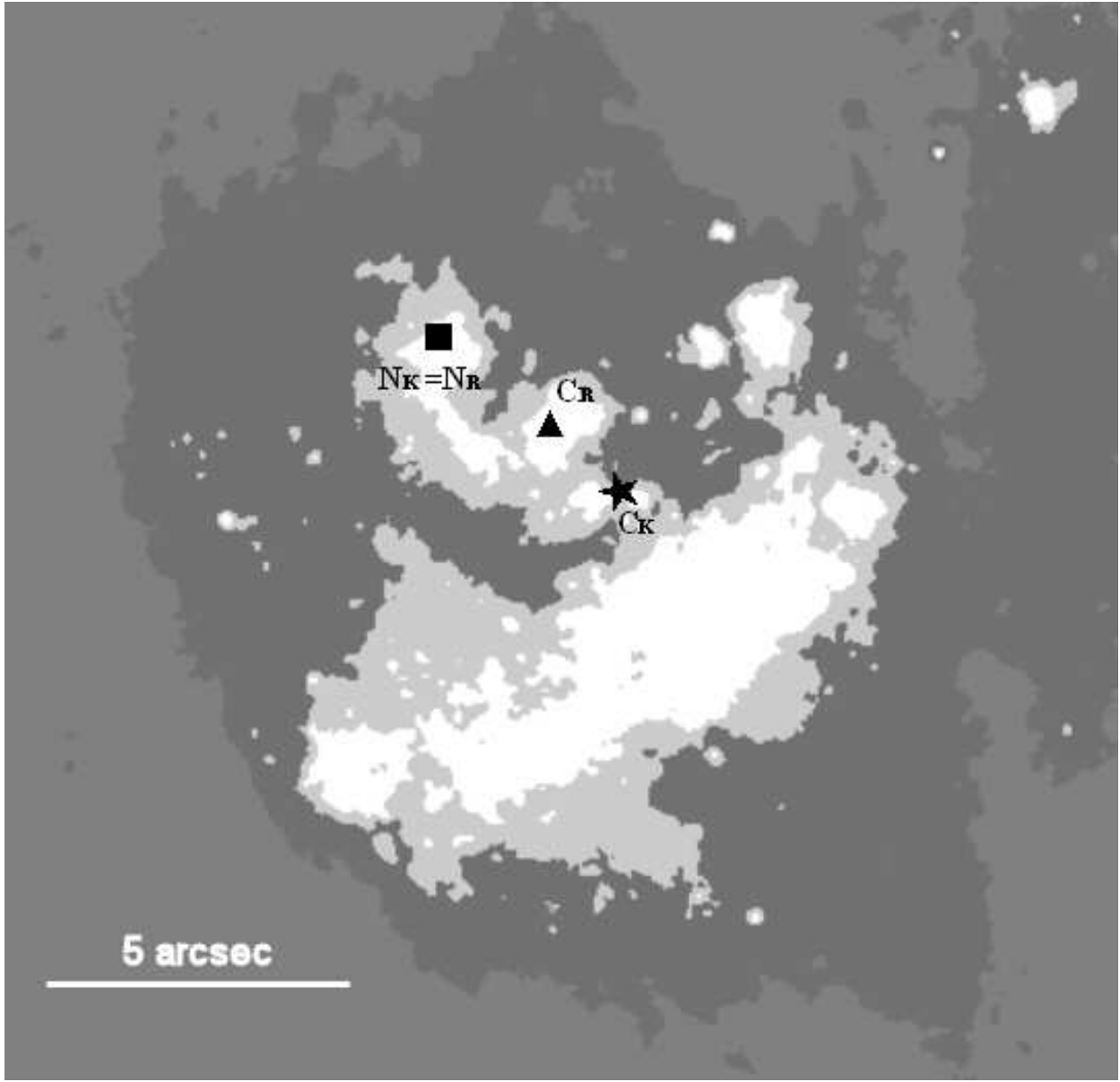}
 \caption{Sketch of NGC 5236 central region. North is at the top
and east is at the left. A square marks the position of the
optical nucleus, which corresponds to the continuum image peak (at
$R$ band, $N_{R}$). A triangle
 marks the position of the red continuum symmetry center ($C_{R}$). A star
 marks the position of the symmetry center of the outer isophotes (bulge) in the
 $K$ band image ($C_{K}$).}
\end{figure}

\begin{figure}
\figurenum{4} \epsscale{1}
 \plotone{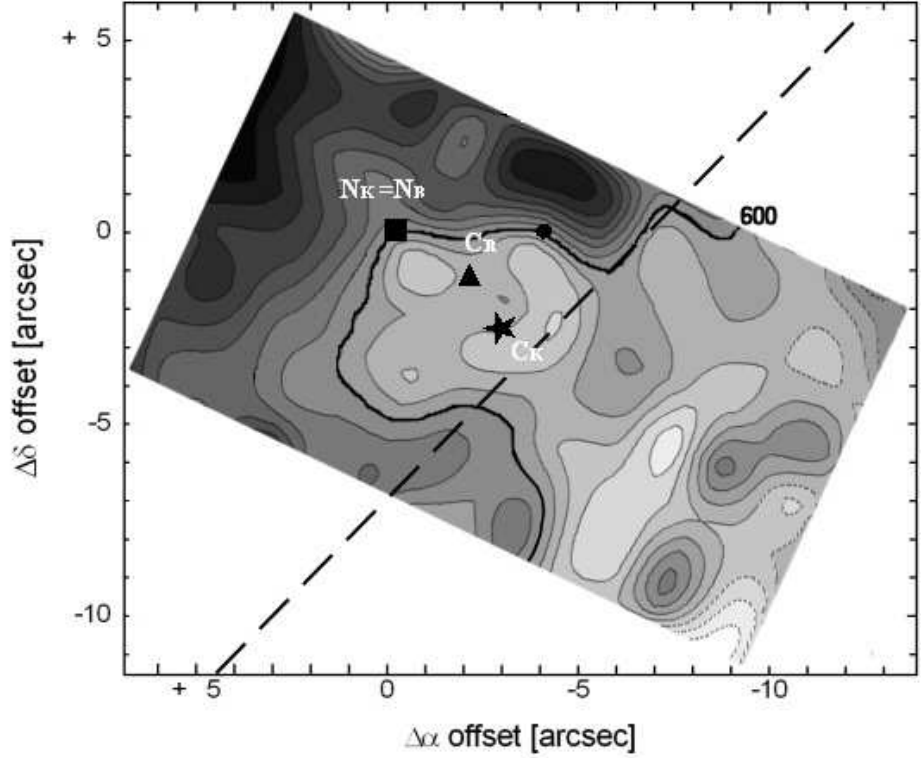}
 \caption{H$\alpha$ radial velocity field. A square indicates the
position of the optical nucleus and a circle marks the position of
the hidden mass concentration. The isovelocities are depicted each
10\,km\,sec$^{-1}$ (2 $\sigma$ uncertainty). The highest and
lowest values are 634\,km\,sec$^{-1}$ and 542\,km\,sec$^{-1}$,
respectively. Dark is blueshift and light is redshift. A thicker
contour corresponds to the systemic velocity. The dotted line
marks the global minor axis at PA $136\degr$. The isovelocity
contours with less robust $S/N$ values have been marked with
dotted lines.}
\end{figure}

\begin{figure}
\figurenum{5} \epsscale{1}
 \plotone{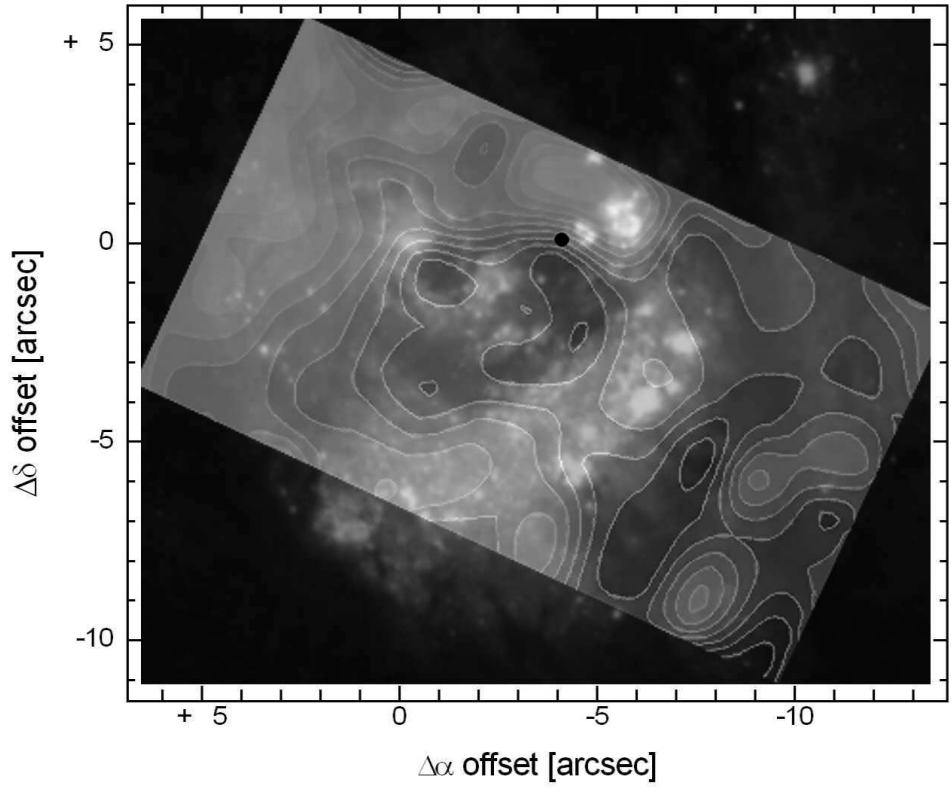}
 \caption{H$\alpha$ radial velocity field + HST grey scale
image. Note the line stretching coincident with the continuum
peak. A circle marks the position of the hidden mass
concentration. Contour levels are the same as in Fig. 4.}
\end{figure}

\begin{figure}
\figurenum{6} \epsscale{1}
 \plotone{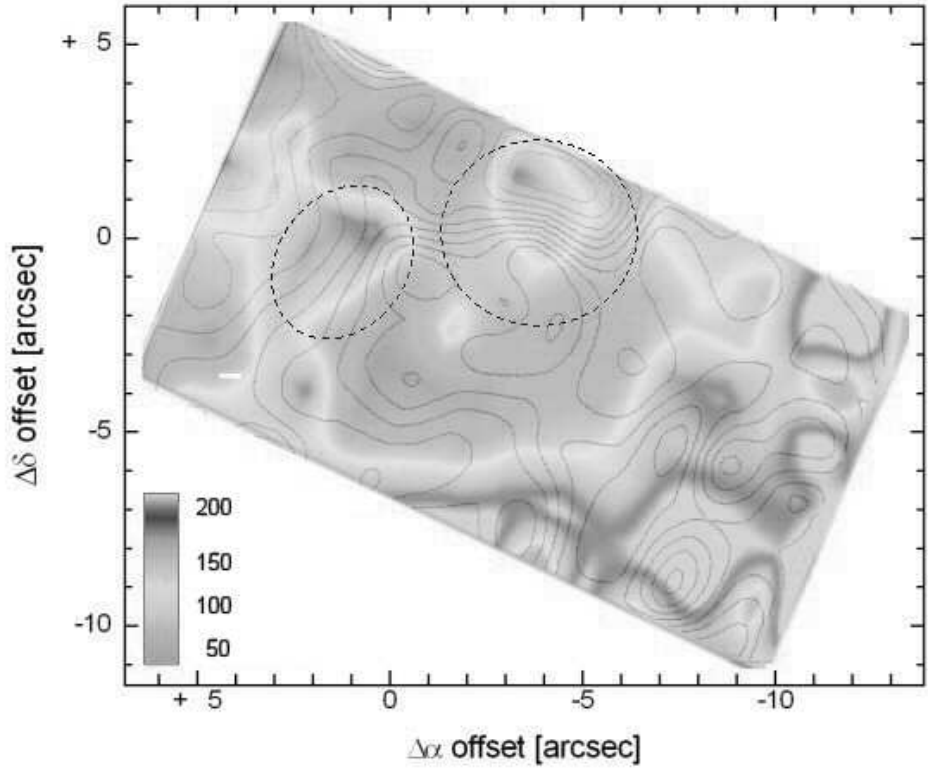}
 \caption{Distribution of radial velocity FWHM, the highest values
are depicted dark. The isolines correspond to the H$\alpha$
velocity field. The darkest zones correspond to FWHM $\sim$ 200 km
sec$^{-1}$. Areas with predominant circular motion are drawn with
dotted lines. In the grey zones with values below 100 km
sec$^{-1}$ (in white) the line widths are not distinct from the
instrumental profile. The inset scale is in km sec$^{-1}$.}
\end{figure}

\begin{figure}
\figurenum{7} \epsscale{1}
 \plotone{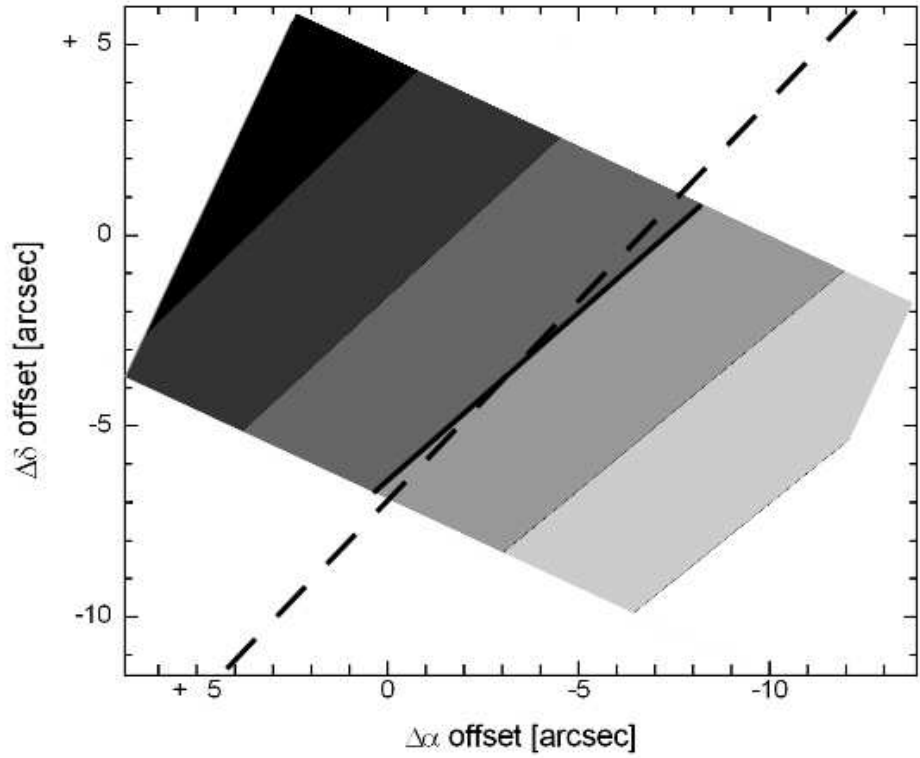}
 \caption{Smoothed ($6\arcsec\times6\arcsec$) radial velocity field. The isovelocities are depicted each
20\,km\,sec$^{-1}$. The minor axis PA of the resultant smoothed
field is 132\degr. A thick contour marks the systemic velocity.
Also drawn with a dotted line is the global CO velocity field
minor axis PA of 136\degr\, (Lundgren et al. 2004).}
\end{figure}

\begin{figure}
\figurenum{8} \epsscale{1}
 \plotone{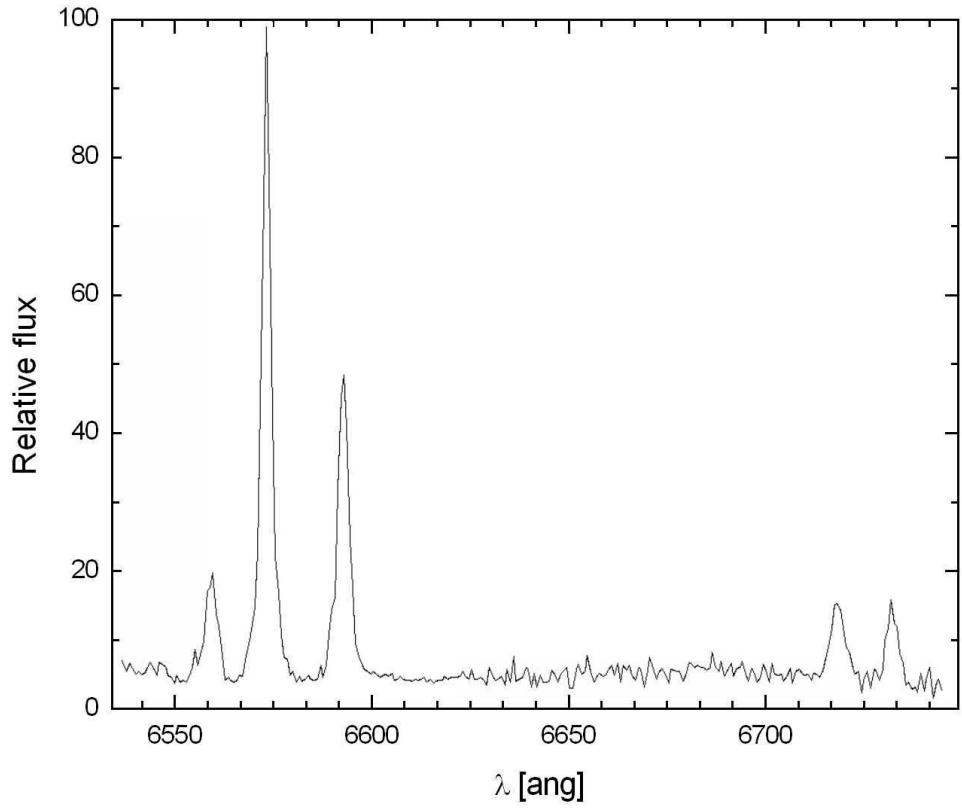}
 \caption{Spectra of a single sampling element from the hidden mass concentration region. Note
that there are not strong asymmetries in the line profiles.}
\end{figure}

\begin{figure}
\figurenum{9} \epsscale{1}
 \plotone{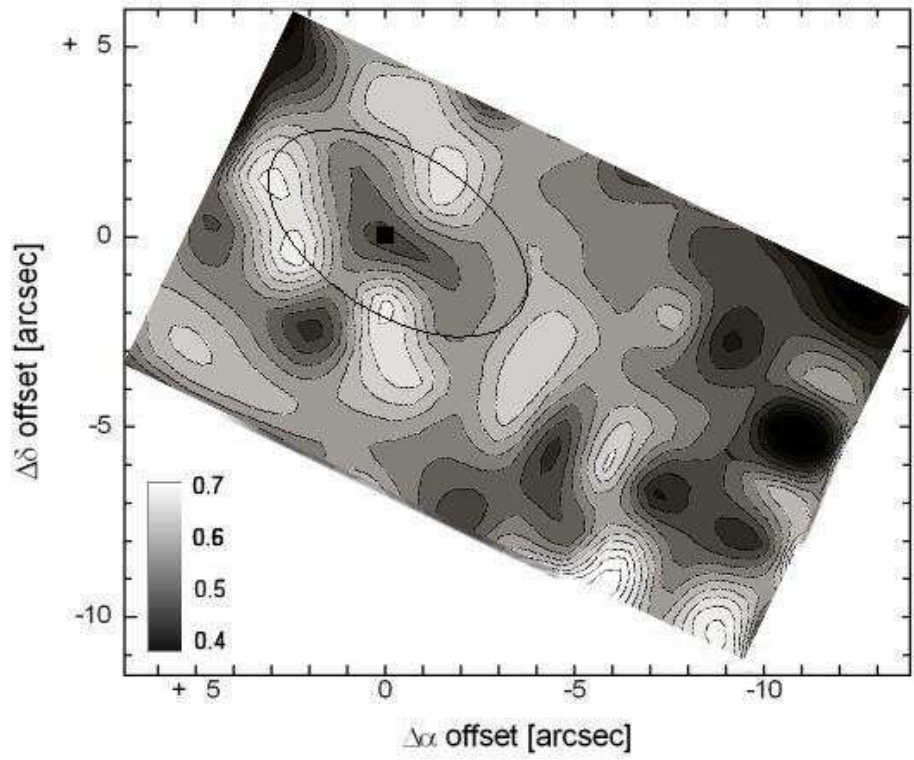}
 \caption{NII/H$\alpha$ ratio map. The peaks of NII/H$\alpha$ ratio
$>$ 0.6 (lightest in the figure) are surrounded by a dark contour.
A square marks the position of the optical nucleus.}
\end{figure}

\begin{figure}
\figurenum{10} \epsscale{1}
 \plotone{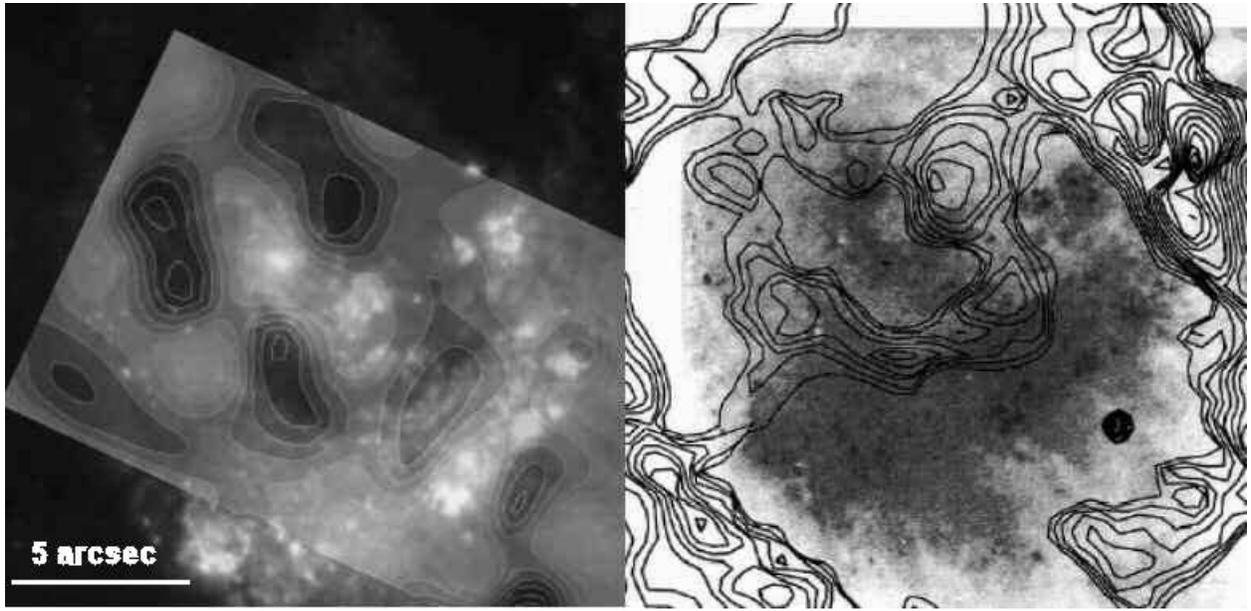}
 \caption{\textit{Left}. NII/H$\alpha$ ratio map superposed to the
grey scale HST image. \textit{Right}. ($J-K$) unsharp-masked
contours overlaid on an HST $V$-band image by Heap (1994) (Figure
5 from Elmegreen et al. 1998).}
\end{figure}

\begin{figure}
\figurenum{11} \epsscale{1}
 \plotone{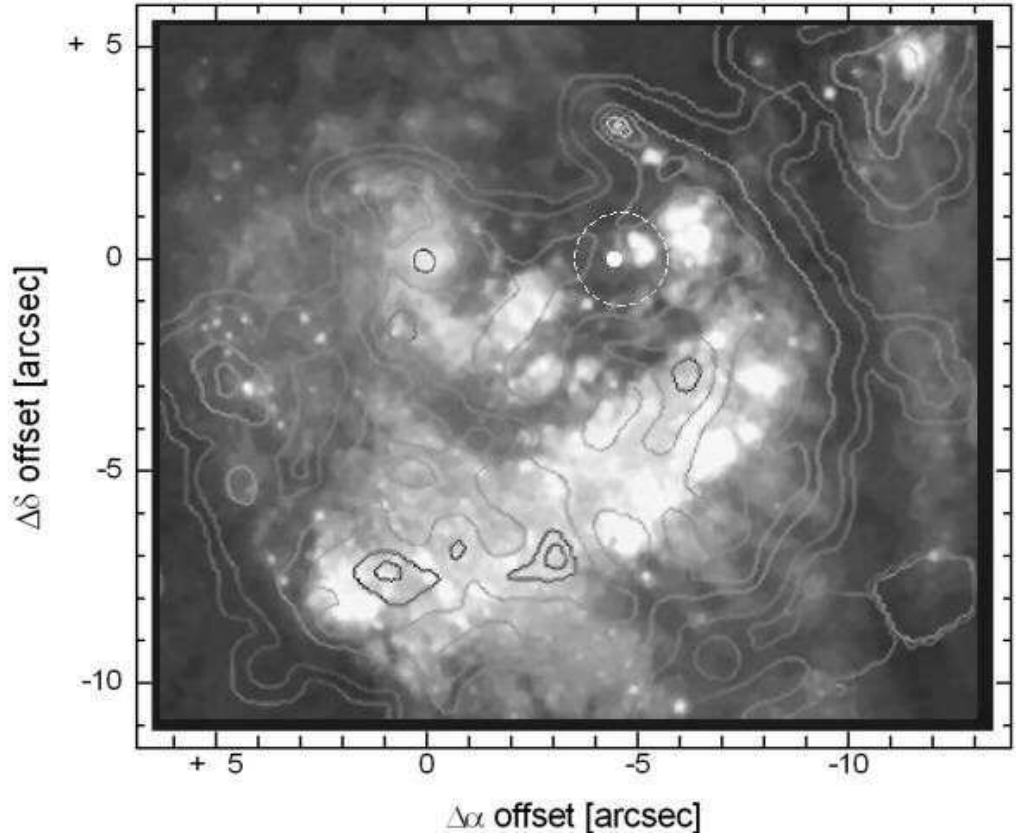}
 \caption{Line emission extinction map
generated from the Pa$\alpha$/H$\alpha$ HST images ratio
(contours) overlaid on a gray scale HST image. Contours values
range from 0.15 at the optical nucleus (dark contours), which
correspond to A$_{V}=0.9$ mag to 2.3 at ($\Delta\alpha$;
$\Delta\delta$)=($-5\arcsec$; $2\farcs5$) (light contours), which
correspond to A$_{V}=6.8$ mag. A white circle marks the position
of the hidden mass concentration and a dotted circle marks the
uncertainty in its location. }
\end{figure}

%TABLES

\begin{deluxetable}{ccc}
\tablecolumns{3} \tablewidth{0pc} \tablecaption{M 83
Characteristics \label{tab0}} \tablehead{ \colhead{Feature}&
\colhead{}& \colhead{Reference}} \startdata
$V_{\odot}$ & 503 km $\sec^{-1}$  &\tablenotemark{1}\\
M  & -20.4  & \tablenotemark{1}\\
Size & $12\arcmin .9 \times 11\arcmin .5$   &\tablenotemark{1,\, 2}\\

B$_T$ & 8.20 mag  &\tablenotemark{2,\, 3}\\
D (h = 0.75)   & 3.7 Mpc & \tablenotemark{3}\\
Kinematic major axis &  PA 46\degr   & \tablenotemark{4}\\
$i$&  24\degr   & \tablenotemark{5}   \\
Inner CN ring radius    & 2.8\arcsec, 51 pc  &\tablenotemark{6}\\
Outer CN ring radius  &  8.6\arcsec, 152 pc  & \tablenotemark{6}\\
Bar length  & 354\arcsec, 6 kpc   & \tablenotemark{7}\\
\enddata
\footnotesize \tablenotetext{1}{Oddone 1999}
\tablenotetext{2}{Measured at $\mu_B = 25\,m{_B}$
arcsec$^{-2}$}\tablenotetext{3}{RC3, de Vaucouleurs et al.
1991}\tablenotetext{4}{Lundgren et al. 2004}
\tablenotetext{5}{Comte 1981} \tablenotetext{6}{Elmegreen et al.
1998} \tablenotetext{7}{DSS image}

\end{deluxetable}

\clearpage

\begin{deluxetable}{ccccccccc} \tablecolumns{9} \tabletypesize{\scriptsize}
\tablecaption{Log of Observations\label{tab1}} \tablehead{
\colhead{ } & \colhead{Date}&\colhead{Exp.Time}&
\colhead{$\lambda_{0}$}& \colhead{$\Delta$$\lambda$}&
\colhead{PA}& \colhead{Pixel scale} & \colhead{Field of view} &
\colhead{Comments}\\ \colhead{} & \colhead{ }& \colhead{[sec]}&
\colhead{[\AA]}& \colhead{[\AA]}& \colhead{[deg]}& \colhead{} &
\colhead{[arcsec]} & \colhead{ }} \startdata

HST &  May 2000 &  600 & 6563 & 21 & 209\degr & 0\farcs049 &
$160\arcsec \times 160\arcsec$ & H$\alpha$ image\\HST &  April
2000 &  160 & 7940 & 1531 & 209\degr & 0\farcs049 & $160\arcsec
\times 160\arcsec$ & Red continuum image\\HST &  May 1999 &  160 &
18739 & 187 & 212\degr & 0\farcs075 & $19\farcs2 \times 19\farcs2$
& Pa$\alpha$ image \\BAlegre &  May 2002 &  3600 & 6600 & 800 &
96\degr &
1\farcs5, 0.78 \AA\ & $12\arcsec \times 21\arcsec$ & Integral Field\\
BAlegre &  May 2002 &  3600 & 6600 & 800 & 96\degr & 1\farcs5,
0.78 \AA\ & $12\arcsec \times 21\arcsec$ & Integral
Field\\\enddata

\tablecomments {BAlegre: Estaci\'on Astrof\'{\i}sica de Bosque
Alegre.}
\end{deluxetable}

\end{document}